\documentclass[journal]{IEEEtran}

\usepackage{cite}
\usepackage{textcase}
\ifCLASSINFOpdf
  \usepackage[pdftex]{graphicx}
\else
\fi
\usepackage{graphicx}
\usepackage{subcaption}
\usepackage{amsmath}
\usepackage{amssymb}
\usepackage{algorithm}
\usepackage{algorithmic}

\newtheorem{assumption}{Assumption}
\newtheorem{theorem}{Theorem}

\begin{document}
%
% paper title
% Titles are generally capitalized except for words such as a, an, and, as,
% at, but, by, for, in, nor, of, on, or, the, to and up, which are usually
% not capitalized unless they are the first or last word of the title.
% Linebreaks \\ can be used within to get better formatting as desired.
% Do not put math or special symbols in the title.
\title{Channel Estimation for Wideband XL-MIMO: A Constrained Deep Unrolling Approach}
%\title{Constrained Deep Unrolling for Wideband XL-MIMO Channel Estimation}
%
%
% author names and IEEE memberships
% note positions of commas and nonbreaking spaces ( ~ ) LaTeX will not break
% a structure at a ~ so this keeps an author's name from being broken across
% two lines.
% use \thanks{} to gain access to the first footnote area
% a separate \thanks must be used for each paragraph as LaTeX2e's \thanks
% was not built to handle multiple paragraphs
%
\author{Peicong~Zheng, Xuantao~Lyu, Ye~Wang,~\IEEEmembership{Member,~IEEE}, and Yi~Gong,~\IEEEmembership{Senior Member,~IEEE}
%\author{Michael~Shell,~\IEEEmembership{Member,~IEEE,}
%        John~Doe,~\IEEEmembership{Fellow,~OSA,}
%        and~Jane~Doe,~\IEEEmembership{Life~Fellow,~IEEE}% <-this % stops a space
\thanks{Peicong Zheng is with Pengcheng Laboratory, Shenzhen 518055, China, and also with the Department of Electrical and Electronic
Engineering, Southern University of Science and Technology, Shenzhen
518055, China.
%(e-mail: 12131052@mail.sustech.edu.cn).
Xuantao Lyu and Ye Wang are with Pengcheng Laboratory, Shenzhen 518055, China.
%(e-mail: \{lvxt, wangy2\}@pcl.ac.cn).
Yi Gong is with the Department of Electrical and Electronic Engineering,
Southern University of Science and Technology, Shenzhen 518055, China
(e-mail: gongy@sustech.edu.cn).}
%\thanks{M. Shell was with the Department
%of Electrical and Computer Engineering, Georgia Institute of Technology, Atlanta,
%GA, 30332 USA e-mail: (see http://www.michaelshell.org/contact.html).}% <-this % stops a space
%\thanks{J. Doe and J. Doe are with Anonymous University.}% <-this % stops a space
%\thanks{Manuscript received April 19, 2005; revised August 26, 2015.
}

\maketitle

% As a general rule, do not put math, special symbols or citations
% in the abstract or keywords.
\begin{abstract}
Extremely large-scale multiple-input multiple-output (XL-MIMO) enables the formation of narrow beams, effectively mitigating path loss in high-frequency communications. 
This capability makes the integration of wideband high-frequency communications and XL-MIMO a key enabler for future 6G networks.
Realizing the full potential of such wideband XL-MIMO systems depends critically on acquiring accurate channel state information. 
However, channel estimation is significantly challenging due to inherent wideband XL-MIMO channel characteristics, including near-field propagation, beam split, and spatial non-stationarity.
To effectively capture these channel characteristics, we formulate channel estimation as a maximum a posteriori problem, which facilitates the use of prior channel knowledge.
We then propose an unrolled proximal gradient descent algorithm with learnable step sizes, which employs a dedicated neural network for proximal mapping.
This design empowers the proposed algorithm to implicitly learn prior channel knowledge directly from data, thereby eliminating the need for explicit regularization functions.
To improve the convergence, we introduce a monotonic descent constraint on the layer-wise estimation error and provide theoretical analyses to characterize the algorithm's convergence behavior.
Simulation results show that the proposed unrolling-based algorithm outperforms the traditional and deep learning-based methods.
\end{abstract}

% Note that keywords are not normally used for peerreview papers.
\begin{IEEEkeywords}
XL-MIMO, channel estimation, deep unrolling, proximal gradient descent.
\end{IEEEkeywords}

% For peer review papers, you can put extra information on the cover
% page as needed:
% \ifCLASSOPTIONpeerreview
% \begin{center} \bfseries EDICS Category: 3-BBND \end{center}
% \fi
%
% For peerreview papers, this IEEEtran command inserts a page break and
% creates the second title. It will be ignored for other modes.
\IEEEpeerreviewmaketitle

\section{Introduction}

Extremely large-scale multiple-input multiple-output (XL-MIMO) is emerging as a promising wireless communication technology, characterized by the deployment of an extremely large number of antenna elements—typically at least an order of magnitude more than those employed in conventional massive MIMO systems \cite{10496996}. 
By substantially increasing the number of antennas, XL-MIMO systems are capable of offering unprecedented spatial degrees of freedom, which can be exploited to significantly enhance wireless link performance.

One of the notable features of XL-MIMO lies in its capability to form ultra-narrow and highly directional beams \cite{10716601}.
Such beamforming capability enables precise spatial focusing of signals, which is especially advantageous in high-frequency wireless communications where severe propagation losses and signal blockages are prevalent, such as in millimeter-wave (mmWave) and terahertz (THz) bands \cite{10494372}.
In addition, the inherently shorter wavelengths at high frequencies facilitate the implementation of compact, high-density antenna arrays.
Consequently, the integration of high-frequency communications, which typically offer wide bandwidths, with XL-MIMO architectures is envisioned as a pivotal enabling technology for future sixth-generation (6G) wireless networks \cite{10054381,10858129}.
This integration is anticipated to deliver ultra-high data rates, massive device connectivity, and enhanced reliability \cite{you2021towards}.
These capabilities are indispensable for supporting a wide range of emerging 6G applications, including immersive extended reality \cite{10183792}, integrated sensing and communication (ISAC) \cite{9606831}, and large-scale Internet of Things (IoT) deployments \cite{10646337}.

Despite these promising prospects, deploying a dedicated radio-frequency (RF) chain for each antenna element in XL-MIMO systems is generally infeasible due to excessive power consumption and hardware cost. 
To facilitate practical deployment, hybrid precoding architectures have been proposed as an effective solution \cite{8030501}.
However, the effectiveness of hybrid precoding relies on the acquisition of accurate channel state information (CSI) \cite{10045774}. 
Since the number of RF chains is much smaller than the number of antenna elements, channel estimation must be performed based on the low-dimensional mixed signals from the RF chains, which leads to excessive pilot overhead as the number of antenna elements increases \cite{9399122}. 
Therefore, efficient channel estimation with reduced pilot overhead is crucial for the practical implementation of wideband XL-MIMO.

\subsection{Prior Work}
In high-frequency communication bands, such as mmWave and THz bands, the 
channel is typically characterized by a limited number of dominant propagation paths \cite{9591285}. 
This is mainly due to the highly directional propagation and relatively few effective scatterers at these frequencies.
As a result, the channel exhibits inherent sparsity, which can be effectively leveraged for efficient channel estimation. 
To exploit this property, compressive sensing (CS) techniques have been widely applied to channel estimation \cite{7842611}. 
In conventional CS-based approaches, the channel is typically sparsified in a certain transform domain through the use of an appropriate dictionary. 
The choice of dictionary plays a crucial role, as it determines the degree to which the channel can be represented sparsely and thus directly affects the performance of the estimation algorithms \cite{8284057}.

In particular, in massive MIMO systems, the discrete Fourier transform (DFT) dictionary is widely used to obtain a sparse channel representation in the angular domain. 
A range of algorithms have been developed to leverage this angular-domain sparsity, including orthogonal matching pursuit (OMP)~\cite{7458188, 8408477}, sparse Bayesian learning (SBL)~\cite{8398433, 8590774}, and approximate message passing (AMP)~\cite{8122055, 8611231}.
In addition to CS-based methods, deep learning-based approaches have been investigated to exploit angular-domain sparsity for channel estimation~\cite{8353153,10178011}.

With the advent of XL-MIMO, the channel characteristics undergo a fundamental change.
Specifically, the substantial increase in the number of antennas and the expansion of the array aperture significantly extend the near-field region \cite{10694053}, such that users and scatterers are more likely to be located within it.
In this near-field region, the electromagnetic wavefronts impinging on the antenna array can no longer be modeled as planar; instead, a spherical wavefront model is more appropriate.
This spherical wave propagation characteristic invalidates the conventional angular-domain sparse representation.
Consequently, the above CS-based channel estimation methods relying on angular sparsity experience significant performance loss in XL-MIMO systems.
To address these challenges, recent studies have proposed constructing polar-domain dictionaries that can capture the inherent joint angle-distance sparsity present in near-field channel~\cite{9693928,10273424,10123941}.
Leveraging the polar-domain sparsity, various CS-based algorithms have been developed for near-field channel estimation, including OMP ~\cite{9693928,10273424},SBL~\cite{10959318}, and compressive sampling matching pursuit (CoSaMP) ~\cite{10418916}.
Besides the aforementioned CS-based approaches, deep learning-based methods have also been extensively explored for XL-MIMO channel estimation~\cite{10044679,10237307,10819470,TWC24}.

In the above channel estimation approaches, the channel is assumed to be spatial stationary, meaning that each multipath component is observable at every antenna in the array.
However, due to the increased array size in XL-MIMO systems, each multipath component may be observed only by a subset of antennas in the array. This phenomenon is known as spatial non-stationarity \cite{9940939}.
This effect can be further characterized by the visibility region, which refers to the specific subset of antennas over which a particular multipath component is observable \cite{9170651}.
The presence of spatial non-stationarity poses a significant challenge for channel estimation.
In~\cite{8949454}, the large-scale antenna array was partitioned into several non-overlapping subarrays, and the channel within each subarray was assumed to be spatial stationary.
Based on this assumption, a subarray-wise OMP algorithm was proposed to perform channel estimation.
The same subarray model was also employed in~\cite{10373799}, where a group time block code based signal extraction scheme was proposed for channel estimation.
While providing useful insights, these works assume that the visibility region of a multipath component is entirely contained within one or more subarrays and covers all their antenna elements.
However, in practice, the visibility region of a multipath component may include only a subset of antenna elements that partially span across subarrays, which is not well captured by the subarray-based models.
Considering the antenna-domain sparsity inherent in spatial non-stationary channel, a structured prior model based on a hidden Markov model was proposed in~\cite{9547795} to capture the characteristics of the visibility region. In addition, the turbo orthogonal approximate message passing algorithm was developed for channel estimation.
In \cite{10780971}, a hierarchical sparse prior was used in the angular domain and a Markov-chain-based prior was imposed in the spatial domain to model the characteristics of spatial non-stationary channel. 
Based on these structured priors, a three-layer generalized approximate message passing algorithm was developed for channel estimation.
In \cite{10715712}, a Markov prior was adopted to model the clustered sparsity of visibility regions, based on which an alternating maximum a posteriori framework was developed for channel estimation.
In~\cite{tang2025channel}, considering dual-wideband effects, a column-wise hierarchical prior was designed to model the structured sparsity of channel for Bayesian-based channel estimation.
%The aforementioned Bayesian approaches have achieved promising results.
%However, these Bayesian methods rely heavily on the accuracy of the hand-crafted priors, which may not fully capture the complex characteristics of XL-MIMO channels.

In addition to the near-field and spatial non-stationarity effects, wideband XL-MIMO systems are subject to frequency-dependent beam split effects \cite{9957130}.
Unlike narrowband systems, where the array steering vectors are approximately frequency-independent, the array steering vectors in wideband systems vary across different subcarriers.
This leads to diverse sparse supports for different subcarriers.
To compensate for the beam split effect, a frequency-dependent dictionary is designed in \cite{10098795}, where the OMP algorithm is employed for channel estimation.
In \cite{cui2023near}, a bilinear pattern prior that captures the joint angle and distance structure of near-field channel is introduced, and a bilinear pattern detection based algorithm is developed for channel estimation.
In \cite{gao2024deep}, a frequency-dependent polar dictionary was developed to facilitate channel estimation, where a unitary AMP–SBL based deep unfolding approach was employed to estimate the channel.
In \cite{10967069}, the frequency-dependent polar dictionary was utilized in the design of a deep learning-based joint learned iterative shrinkage thresholding algorithm with partial weight-coupling for wideband channel estimation.
In \cite{10486834}, a wideband redundant dictionary was proposed to enable a knowledge-driven mixed-field wideband channel estimation framework.
In \cite{10606003}, a Bernoulli-Gaussian prior model was employed to capture beam-delay domain sparsity, and hybrid message passing algorithms were developed for wideband channel estimation.

As discussed above, existing works have primarily addressed either the joint modeling of near-field propagation and spatial non-stationarity (typically under narrowband conditions) or the joint modeling of near-field and wideband beam split effects (often under the assumption of spatial stationarity). 
While the above prior-based and dictionary-based approaches have demonstrated promising performance, they fundamentally rely on handcrafted priors and dictionaries. 
However, the interplay among near-field propagation, spatial non-stationarity, and beam split effects gives rise to highly complex channel behaviors in wideband XL-MIMO systems, making it intractable to design handcrafted priors or dictionaries that can accurately represent such intricacies. 
Consequently, these methods may fail to fully capture the underlying channel structure, leading to degraded estimation performance.
In addition, the aforementioned works are primarily based on uniform linear array (ULA) configurations. 
When extended to more complicated uniform planar array (UPA) configurations, the coupling among the two-dimensional azimuth-elevation angle pair, distance parameters, spatial non-stationarity, and beam split effects renders it even more challenging to design suitable priors or dictionaries for accurate channel estimation. %vr不要缩小
These limitations highlight the need for a more flexible approach that can learn the underlying channel structure.
\subsection{Our Contributions}
In light of the aforementioned challenges and limitations, the main contributions of this paper are as follows:
\begin{itemize}
\item We develop a  wideband XL-MIMO channel model for both ULA and UPA configurations that characterizes the near-field propagation, spatial non-stationarity, and wideband beam split effects.
To overcome the challenges associated with dictionary-based approaches, we formulate the channel estimation problem as a maximum a posteriori (MAP) optimization problem.
This MAP formulation naturally incorporates the channel's characteristics through the prior knowledge of channel without requiring sparse representation in a predefined dictionary.
\item We introduce the proximal gradient descent (PGD) algorithm as an effective solver for the MAP problem.
However, designing appropriate regularization functions that accurately capture the characteristics of channel presents significant challenges.
This directly affects PGD implementation since each iteration requires computing the proximal operator, which depends on the regularization function.
To address this limitation, we propose an unrolled PGD network that transforms PGD into a layer-wise network by introducing learnable parameters and a neural network-based proximal operator.
Consequently, this design eliminates the need for explicit regularization functions design while implicitly learning the channel's characteristics.
\item We enhance the proposed unrolled PGD network by incorporating a monotonic descent constraint across layers to improve convergence behavior and introducing noise injection during training.
The constraint ensures that each layer's output progressively reduces the distance to the ground truth channel.
We solve this constrained learning problem using a primal-dual training algorithm and establish theoretical convergence guarantees.
Simulations with ULA and UPA configurations demonstrate that our approach consistently outperforms traditional and  deep learning-based methods.
%传统的基于字典的压缩感知方法依赖于信道在特定基底(如极坐标域)下具有稀疏性
%然而，在宽带XL-MIMO系统中，波束分裂效应破坏了跨子载波的共同支撑结构，空间非平稳性导致功率分散，这些都严重损害了信道的稀疏性
%MAP方法的优势在于：
%通过先验分布p(h)直接建模信道特性，不需要依赖信道在特定域中的稀疏表示
%正则化项R(h)可以灵活设计，以捕获信道的结构信息
%避开了对稀疏性的依赖，从而能够更好地处理波束分裂和空间非平稳性带来的挑战
\end{itemize}

The remainder of this paper is organized as follows. 
Section II presents the system and channel model, formulating channel estimation as a MAP problem. 
Section III introduces the unrolled PGD network. 
Section IV details the proposed constrained unrolled PGD network.
Section V provides simulation results, and Section VI concludes the paper.
\section{System Model and Problem Formulation}
\subsection{Signal model}
\begin{figure*}[htbp]
        \centering
	{\includegraphics[scale=0.65]{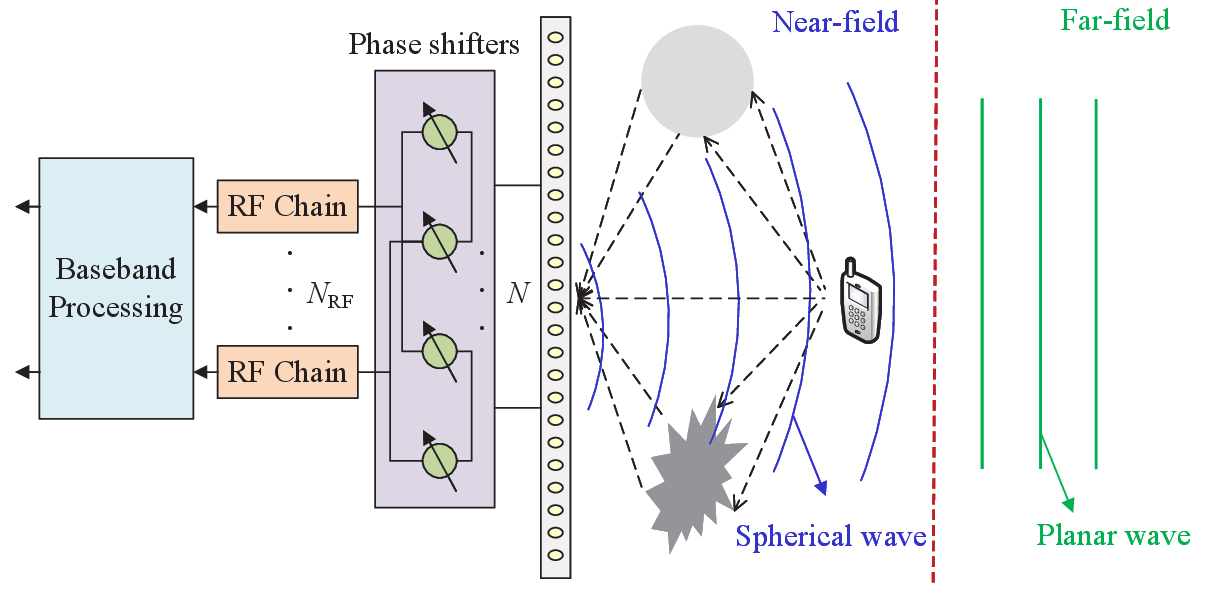}}
	\caption{System diagram of a hybrid precoding architecture showing near-field and far-field propagation.}
        \label{sys}
\end{figure*}

In this paper, we investigate an uplink time-division-duplex (TDD) based wideband XL-MIMO system. The base station (BS) is equipped with a fully-connected hybrid precoding architecture comprising $N_{\mathrm{RF}}$ radio frequency (RF) chains and an $N$-antenna array to serve single-antenna users, as illustrated in Fig.~\ref{sys}.
The system operates with orthogonal frequency division multiplexing (OFDM) with $M$ subcarriers. 
The $m$-th subcarrier frequency is given by $f_m = f_c + \frac{2m - M - 1}{2}B$, where $f_c$ denotes the center frequency and $B$ is the bandwidth.

For the hybrid precoding architecture to achieve its full potential, accurate CSI is crucial. 
By exploiting the channel reciprocity in TDD systems, we adopt an uplink channel estimation scheme where users transmit orthogonal pilots to the BS.
Due to pilot orthogonality, channel estimation for each user can be performed independently. 
Omitting the user index for notational simplicity, the received signal $\mathbf{y}_{m, p} \in \mathbb{C}^{N_{\mathrm{RF}} \times 1}$ at the BS during the  $p$-th pilot transmission slot at the $m$-th subcarrier can be expressed as
\begin{equation}
\mathbf{y}_{m, p}=\mathbf{A}_p \mathbf{h}_m s_{m, p}+\ \mathbf{n}_{m, p},
\end{equation}
where $\mathbf{A}_p \in \mathbb{C}^{N_{\mathrm{RF}} \times N}$ denotes the combining matrix, $\mathbf{h}_m \in \mathbb{C}^{N \times 1}$ represents the $m$-th subchannel, $s_{m, p}$ denotes the pilot symbol, and $\mathbf{n}_{m, p} \in \mathbb{C}^{N_\mathrm{RF} \times 1}$ denotes the complex Gaussian noise following the distribution $\mathcal{C N}\left(0, \sigma^2 \mathbf{I}_{N_{\mathrm{RF}}}\right)$ with the noise power $\sigma^2$. Since the pilots are known to the BS, we set $s_{m,p} = 1$ without loss of generality.

By concatenating the received signals over $P$ pilot transmission slots into $\mathbf{y}_m=\left[\mathbf{y}_{m, 1}^T, \ldots, \mathbf{y}_{m, P}^T\right]^T \in \mathbb{C}^{P N_{\mathrm{RF}} \times 1}$, we can express the received signal at the $m$-th subcarrier as
\begin{equation}
    \mathbf{y}_m=\mathbf{A} \mathbf{h}_m+\mathbf{n}_m,
\end{equation}
where $\mathbf{A}=\left[\mathbf{A}_1^T, \ldots, \mathbf{A}_P^T\right]^T \in \mathbb{C}^{P N_{\mathrm{RF}} \times N}$ is the stacked combining matrix, and $\mathbf{n}_m=\left[\mathbf{n}_{m, 1}^T , \ldots, \mathbf{n}_{m, P}^T\right]^T \in$ $\mathbb{C}^{P N_{\mathrm{RF}} \times 1}$ is the equivalent noise.
\subsection{Channel Model}
In XL-MIMO systems, the unprecedented array aperture introduces unique channel characteristics that fundamentally differ from those in conventional massive MIMO systems. 
Unlike traditional channel models that rely on the far-field assumption, where electromagnetic waves are typically modeled as plane waves, XL-MIMO channel exhibit three distinctive features that demand careful consideration: 
\begin{itemize}
    \item Near-field propagation:
    The wireless channel exhibits distinct characteristics in the far-field and near-field regions. The boundary between these regions is typically determined by the Rayleigh distance, defined as  $r_{RD} = 2D^2/\lambda$, where $D$ is the array aperture and $\lambda$ is the wavelength. 
    In conventional massive MIMO systems, due to the limited array dimensions, the Rayleigh distance is approximately 5 meters, placing most users in the far-field region.
    The wavefront in the far-field region can be simply modeled as a planar wavefront.
    Thus, the array steering vectors depend solely on the angle of departure/arrival (AoD/AoA).
    However, XL-MIMO systems introduce significantly larger array apertures, extending the Rayleigh distance to hundreds of meters.
    For example, a 0.4-meter array operating at 100 GHz results in a Rayleigh distance of about 107 meters. 
    In this extended near-field region, the spherical wavefront model becomes necessary, in which the array steering vectors are determined by both the angle and the distance between the array and scatterers.
    \item Frequency-dependent beam split: 
    In conventional narrowband systems where $f_m \approx f_c$, the array steering vectors can be approximated as frequency-independent, which leads to a common sparse support structure across all subcarriers. 
    Such common sparsity facilitates joint channel estimation over subcarriers. 
    However, in wideband XL-MIMO systems, the array steering vectors exhibit strong frequency dependence due to the substantial offsets of subcarrier frequencies from the center frequency. 
    For example, consider a system with $M=256$ subcarriers, a center frequency of $f_c = 100$ GHz, and a bandwidth of $B = 10$ GHz. 
    In this case, the first and last subcarriers, $f_1 = 95$ GHz and $f_M = 105$ GHz, are each offset from the center frequency by 5 GHz in opposite directions, which leads to pronounced differences in the array steering vectors.
    %This phenomenon, known as the beam split effect, poses significant challenges for channel estimation, as the shared sparsity can no longer be jointly exploited across subcarriers.
    \item Spatial non-stationarity: In conventional MIMO systems, the whole array is visible to all multipath components. 
    However, in XL-MIMO systems, due to the extremely large array aperture, different regions of the array may be visible to different multipath components.
    This difference in visible regions leads to the spatial non-stationarity of the channel.
\end{itemize}

To capture these characteristics, we establish channel models for two antenna configurations at the BS: uniform linear array and uniform planar array.
\subsubsection{ULA Configuration}
Consider a ULA with $N$ antennas deployed along the $x$-axis with antenna spacing $d=\lambda_c/2$, where $\lambda_c$ is the wavelength at the center frequency.
For the ULA configuration, the channel response at the $m$-th subcarrier can be expressed as
\begin{equation}
    \mathbf{h}_m= \sum_{l=1}^{L} \alpha_{l,m} \mathrm{e}^{-\mathrm{j} \frac{2 \pi}{\lambda_m} r_l}\mathbf{b}\left(r_{l}, \theta_{l},f_m\right) \odot \mathbf{q}\left(\mathrm{VR}_{l}\right),
\end{equation}
where $L$ denotes the number of paths, $\alpha_{l,m}$ represents the complex path gain of the $l$-th path, $\theta_l$ is the angle of arrival (AoA), and $r_l$ represents the distance between the scatterer and the array center. 

\begin{figure}[htbp]
\centering
\includegraphics[scale=0.975]{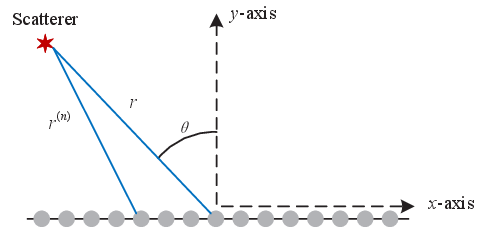}
\caption{Scatterer distance and angles relative to the ULA.}
\label{ULA}
\end{figure}

Under the spherical wavefront assumption in the near-field region, the steering vector $\mathbf{b}(r, \theta,f)$ can be formulated as
\begin{equation}
\mathbf{b}(r, \theta,f)=\frac{1}{\sqrt{N}}\left[e^{-j \frac{2\pi f}{c}(r^{(1)}-r)}, \cdots, e^{-j \frac{2\pi f}{c}(r^{(N)}-r)}\right]^T,
\end{equation}
where $c$ is the speed of light and the relevant geometry is shown in Fig.~\ref{ULA}.
For the $l$-th path, the distance $r^{(n)}$ between the scatterer and the $n$-th BS antenna can be calculated as
\begin{equation}
r_{l}^{(n)}=\sqrt{r_{\ell}^2-2 r_{l} \delta_n d \theta_{l}+\delta_n^2 d^2}
\end{equation}
with $\delta_n=\frac{2 n-N-1}{2}$.

The spatial non-stationary characteristic emerges in XL-MIMO channels due to the significantly enlarged array aperture, where each multipath component may be observed by only a subset of the antenna array elements, rather than by the entire array.
The visibility relationship can be mathematically characterized through $\mathrm{VR}_l$, defined as the visibility region corresponding to the $l$-th path. 
This spatial selectivity is captured by the binary visibility mask $\mathbf{q}(\mathrm{VR}_l)$, defined as
\begin{equation}
\left[\mathbf{q}\left(\mathrm{VR}_l\right)\right]_n= \begin{cases}1, & n \in \mathrm{VR}_l \\ 0, & n \notin \mathrm{VR}_l\end{cases}
\label{mask}
\end{equation}

\subsubsection{UPA Configuration} 
\begin{figure}[htbp]
\centering
\includegraphics[width=\columnwidth]{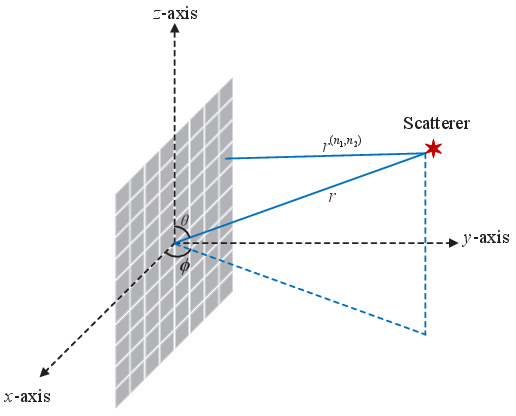}
\caption{Scatterer distance and angles relative to the UPA.}
\label{UPA}
\end{figure}

For a UPA with $N = N_1 \times N_2$ antennas, where $N_1$ and $N_2$ represent the number of elements along the $x$- and $z$-axis respectively, the channel response at the $m$-th subcarrier can be expressed as
\begin{equation}
    \mathbf{h}_m= \sum_{l=1}^{L} \alpha_{l,m} \mathrm{e}^{-\mathrm{j} \frac{2 \pi}{\lambda_m} r_l}\mathbf{b}\left(r_{l}, \phi_{l}, \theta_{l},f_m\right) \odot \mathbf{q}\left(\mathrm{VR}_{l}\right),
\end{equation}
where $\phi_l$ and $\theta_l$ represent the azimuth and elevation angles, respectively. The relevant geometry is shown in Fig.~\ref{UPA}.

Considering the spherical wavefront in the near-field region, the UPA steering vector $\mathbf{b}(r, \phi, \theta,f)$ can be formulated as
\begin{equation}
\begin{array}{r}
\mathbf{b}(r, \phi, \theta,f)=\frac{1}{\sqrt{N}}\left[e^{-j \frac{2\pi f}{c}(r^{(1,1)}-r)}, \cdots,\right. \\
\left.e^{-j \frac{2\pi f}{c}(r^{({N}_1,{N}_2)}-r)}\right]^T.
\end{array}
\end{equation}
For the $l$-th path, the distance $r_l^{(n_1,n_2)}$ between the scatterer and the $(n_1,n_2)$-th antenna element can be calculated as
\begin{small}
\begin{equation}
\begin{aligned}
&r_l^{\left(n_1, n_2\right)} = \\
& \sqrt{\left(r_l \sin \theta_l \cos \phi_l\right)^2+\left(r_l \sin \theta_l \sin \phi_l-\delta_{n_1} d\right)^2+\left(r_l \cos \theta_l-\delta_{n_2} d\right)^2}
\end{aligned}
\end{equation}
\end{small} with $\delta_{n_1}=\frac{2 n_1-N_1-1}{2}$ and $\delta_{n_2 }=\frac{2 n_2-N_2-1}{2}$.

Similar to the ULA case, $\mathbf{q}\left(\mathrm{VR}_{l}\right)$ represents the binary visibility mask that captures the spatial non-stationarity characteristics, as defined in \eqref{mask}, where each element indicates whether the corresponding antenna element can observe the $l$-th path, with $1$ denoting visibility and $0$ denoting invisibility.

\subsection{Problem formulation}
CS-based channel estimation approaches represent the near-field channel in the polar domain to exploit its sparse structure. 
However, the beam split effect invalidates the common support assumption across subcarriers in the polar domain, making joint estimation across subcarriers infeasible.
Moreover, since the polar-domain dictionary is derived under the spatial stationarity assumption, the spatial non-stationarity leads to power dispersion, destroying the channel sparsity that CS methods rely on.
% 单栏图片

\begin{figure}[!t]
    \centering
    \includegraphics[width=3.4in]{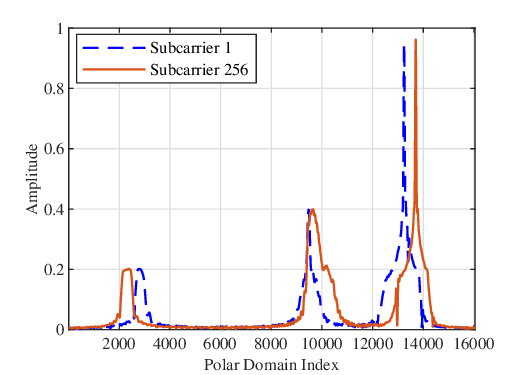}
    \caption{Impact of beam split and spatial non-stationarity on polar-domain channel representation.}
    \label{fig:single}
\end{figure}

To illustrate these challenges, we present a numerical example in Fig. \ref{fig:single} for a ULA with $N=512$ antennas, $M=256$ subcarriers, $f_c=100$ GHz, and $B=10$ GHz.
Three paths are considered with distances $[10\text{m}, 20\text{m}, 50\text{m}]$, angles $[-\pi/6, \pi/24, \pi/6]$, and visibility ratios $[25\%, 50\%, 100\%]$, respectively.
The beam split effect can be observed from the third path (with $100\%$ visibility ratio), where the peak locations in the polar domain differ between subcarrier $1$ and subcarrier $256$.
This frequency-dependent shift in support invalidates the common sparsity assumption across subcarriers, making joint processing challenging. 
The impact of spatial non-stationarity on channel sparsity is demonstrated by comparing paths with different visibility ratios. 
The first path with $25\%$ visibility exhibits severe power dispersion with a lower and broader peak, while the second path with $50\%$ visibility shows moderate dispersion. In contrast, the third path with full visibility maintains good sparsity with a sharp peak. 
This degradation in sparsity, particularly pronounced for paths with limited visibility, fundamentally challenges conventional CS-based estimation methods that rely on the channel's sparse nature in the polar domain.

Given these fundamental challenges, we approach the channel estimation problem by formulating it as a maximum a posteriori estimation problem. 
According to the Bayesian inference framework, the MAP estimation of $\mathbf{h}$ given the $\mathbf{y}$ and  $\mathbf{A}$ can be formulated as
\begin{equation}
\begin{aligned}
\hat{\mathbf{h}} &= \arg\max_{\mathbf{h}} p(\mathbf{h}|\mathbf{y}) =\arg\max_{\mathbf{h}} \frac{p(\mathbf{y}|\mathbf{h})p(\mathbf{h})}{p(\mathbf{y})} \\
&= \arg\max_{\mathbf{h}} p(\mathbf{y}|\mathbf{h})p(\mathbf{h}),
\end{aligned}
\end{equation}
where $p(\mathbf{y}|\mathbf{h})$ is the likelihood function, and $p(\mathbf{h})$ represents the prior distribution of the channel capturing its characteristics. Taking the negative logarithm, the MAP estimation is equivalent to
\begin{equation}
\begin{aligned}
\hat{\mathbf{h}} &= \arg\min_{\mathbf{h}} {-\log p(\mathbf{y}|\mathbf{h}) - \log p(\mathbf{h})}\\
&= \arg\min_{\mathbf{h}} \frac{1}{2}\|\mathbf{y} - \mathbf{A}\mathbf{h}\|_2^2+\rho R(\mathbf{h}),
\label{MAP}
\end{aligned}
\end{equation}
where the first term $\|\mathbf{y} - \mathbf{A}\mathbf{h}\|_2^2$ corresponds to $-\log p(\mathbf{y}|\mathbf{h})$ and serves as the data fidelity term. 
The regularization term $\rho R(\mathbf{h})$ with $R(\mathbf{h})=-\log p(\mathbf{h})$ incorporates the channel characteristics and $\rho$ controls the trade-off between data fidelity and regularization.
\section{Unrolled PGD}
\subsection{PGD}
In this section, we first introduce the proximal gradient descent algorithm \cite{parikh2014proximal} as a fundamental algorithm to solving the MAP estimation problem formulated in \eqref{MAP}. 
PGD is particularly effective in handling optimization problems with non-smooth regularization functions. 
The PGD iteration proceeds as follows

\begin{subequations}
\begin{align}
& \mathbf{z}_{t} = \mathbf{h}_{t-1} - \alpha \mathbf{A}^T(\mathbf{A}\mathbf{h}_{t-1} - \mathbf{y}) \label{zt} \\
& \mathbf{h}_{t} = \text{prox}_{\alpha\rho R}(\mathbf{z}_{t}),
\label{ht}
\end{align}
\end{subequations}
where the first step in \eqref{zt} performs gradient descent on the data fidelity term with step size $\alpha$. 
The proximal step in \eqref{ht} involves the proximal operator $\text{prox}_{\alpha\rho R}(\cdot)$, defined as
\begin{equation}
\text{prox}_{\alpha\rho R}(\mathbf{z}) = \arg\min_{\mathbf{h}} {\frac{1}{2}\|\mathbf{h}-\mathbf{z}\|_2^2 + \alpha\rho R(\mathbf{h})}.
\end{equation}
\begin{figure*}[htbp]
        \centering
	{\includegraphics[scale=0.85]{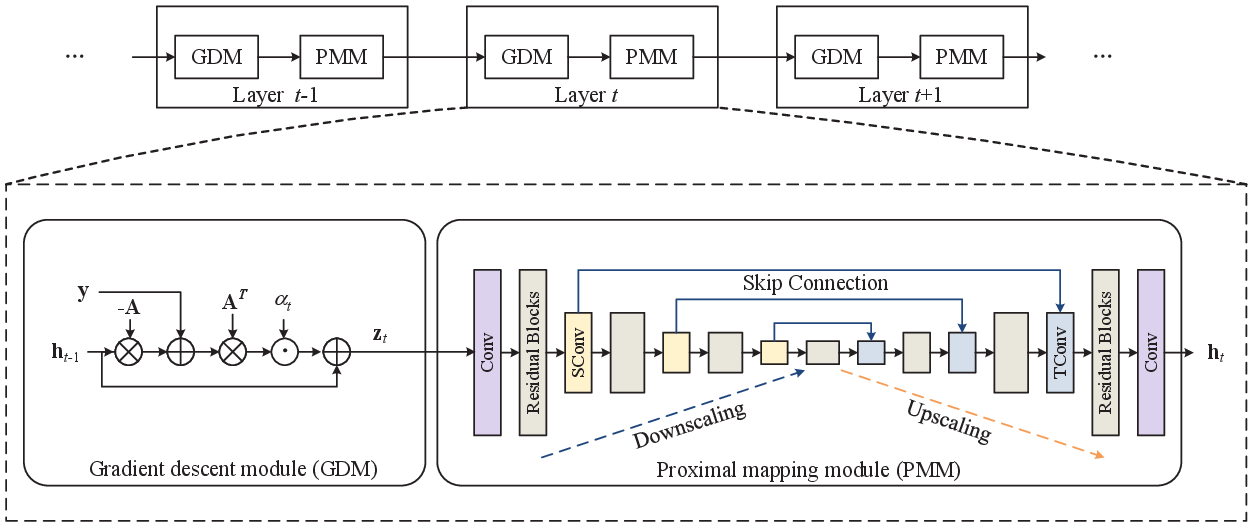}}
	\caption{Architecture of the proposed unrolled proximal gradient descent network.}
        \label{CNN_CDL}
\end{figure*}

The proximal step plays a crucial role in incorporating prior knowledge of the channel characteristics, as encoded in the regularization function $R(\mathbf{h})$.
The choice of $R$ is critical: if it is too simple, it may fail to capture the essential properties of the channel, leading to suboptimal solutions.
Conversely, if $R$ is overly complex, the increased difficulty of solving the proximal operator can make the optimization process computationally intractable. 
Moreover, in practice, deriving an explicit form of $R$ that accurately models the channel characteristics is challenging, particularly in scenarios where prior information is limited or the channel exhibits complex behavior such as near-field propagation , frequency-dependent beam split, and spatial non-stationary.
%The key points to emphasize:
%In PGD, the proximal operator directly depends on the regularization function
%Computing the proximal operator requires solving a subproblem at each iteration
%For complex channel models, finding closed-form expressions for the proximal operator is often impossible
%Without an efficient way to compute the proximal operator, PGD becomes computationally prohibitive
%The regularization function needs to capture complex channel characteristics while maintaining computational tractability
\subsection{Unrolled PGD}
To overcome these limitations, we propose a deep unrolling approach that transforms the iterative PGD algorithm into a learnable architecture.
Specifically, we unroll the PGD algorithm for a fixed number of $T$ iterations and introduce two key modifications: (1) learning the step size in the gradient descent step, and (2) employing a dedicated neural network to perform proximal mapping.

In the standard PGD algorithm, the step size $\alpha$ in the gradient step is fixed and manually selected, which limits flexibility. 
%To address this limitation, we introduce a learnable step size $\alpha_t$ for each iteration.
To address this limitation, a learnable step size $\alpha_t$ is introduced at each iteration.
The gradient step is given by
\begin{equation}
\mathbf{z}_{t} = \mathbf{h}_{t-1} - \alpha_t \mathbf{A}^T(\mathbf{A}\mathbf{h}_{t-1} - \mathbf{y}),
\label{eq:learnable_step}
\end{equation}
where $\alpha_t$ is specific to each iteration and optimized during training. 
The proximal step in PGD involves computing the proximal operator, which depends on the regularization function $R(\mathbf{h})$. 
To avoid explicitly defining $R(\mathbf{h})$, we replace the proximal step with a trainable neural network $\mathcal{N}_{\theta_t}$, specific to each iteration $t$. The proximal update is reformulated as
\begin{equation}
\mathbf{h}_{t} = \mathcal{N}_{\theta_t}(\mathbf{z}_{t}),
\label{eq:prox_net}
\end{equation}
where $\mathcal{N}_{\theta_t}$, parameterized by $\theta_t$, maps the intermediate variable $\mathbf{z}_{t}$ to the updated estimate $\mathbf{h}_{t}$. 
This data-driven approach eliminates the need for handcrafted regularization functions by allowing the network to implicitly learn prior information from data, enabling the model to flexibly adapt to complex channel characteristics.

The network $\mathcal{N}_{\theta_t}$ adopts a U-Net architecture that integrates residual blocks into its multi-scale structure. 
As illustrated in Fig.~\ref{CNN_CDL}, the architecture consists of an encoder-decoder structure with identity skip connections across four scales. 
The encoder downsamples the input using $2 \times 2$ strided convolutions (SConv), while the decoder upsamples features using $2 \times 2$ transposed convolutions (TConv).
The number of feature channel at each scale increases progressively from 64 to 128, 256, and 512 channels, respectively. 
Following the design principles in \cite{lim2017enhanced}, no activation functions are applied after the first and last convolutional layers, as well as the strided convolution and transposed convolution layers. 
Moreover, each residual block contains only a single ReLU activation function.
This combination of U-Net and residual blocks allows $\mathcal{N}_{\theta_t}$ to flexibly capture both local and global features.

Building upon the unrolled architecture described in \eqref{eq:learnable_step} and \eqref{eq:prox_net}, the unrolled PGD network can be represented as $\mathbf{\Phi}(\mathbf{y} ;\mathbf{W})$ with trainable parameters $\mathbf{W} = \left[{\theta_t, \alpha_t}\right]_{t=1}^T$.
The training objective is formulated as
\begin{equation}
{\underset{\mathbf{W}}{\operatorname{argmin}}}\
\mathbb{E}\left[\ell\left(\mathbf{\Phi}(\mathbf{y} ; \mathbf{W}), \mathbf{h}\right)\right],
\label{eq:training}
\end{equation}
where $\ell(\cdot,\cdot)$ denotes $\ell_2$ measuring the estimation error between the estimated and ground truth channel.

\section{Constrained Unrolled PGD}
\subsection{Monotonic Descent Constraint}
To enhance the convergence behavior of the unrolled PGD network, we impose a monotonic descent constraint $\mathcal{C}(\cdot, \cdot)$ on the distance between the network output and the ground truth channel across layers. Specifically, the output of the $t$-th unrolled layer, $\mathbf{h}_t$, is constrained by
%\begin{equation}
%\mathbb{E}\left[\mathcal{C}\left(\mathbf{h}_t, \mathbf{h}_{t-1}\right)\right]=\mathbb{E}\left[\left\|\mathbf{h}_t-\mathbf{h}\right\|_2-(1-\epsilon)\left\|\mathbf{h}_{t-1}-\mathbf{h}\right\|_2\right] \leq 0,
%\label{cons}
%\end{equation}
\begin{equation}
\mathcal{C}\left(\mathbf{h}_t, \mathbf{h}_{t-1}\right)=\left\|\mathbf{h}_t-\mathbf{h}\right\|_2-(1-\epsilon)\left\|\mathbf{h}_{t-1}-\mathbf{h}\right\|_2 \leq 0,
\label{cons}
\end{equation}
where $\epsilon > 0$ is a contraction factor that promotes strict descent. 
This constraint encourages the unrolled PGD network to produce a sequence of outputs that progressively approach the optimal solution.
Integrating this constraint into the training objective in \eqref{eq:training} results in the following constrained learning process
\begin{equation}
\begin{aligned}
P &= \underset{\mathbf{W}}{\operatorname{min}}  \quad \mathbb{E}\left[\ell\left(\boldsymbol{\Phi}(\mathbf{y} ;  \mathbf{W}), \mathbf{h}\right)\right] \\
\text{s.t.} & \quad \mathbb{E}\left[\mathcal{C}\left(\mathbf{h}_t, \mathbf{h}_{t-1}\right)\right] \leq 0, \, t \in T.
\label{CO}
\end{aligned}
\end{equation}

To further enhance the robustness of the unrolled PGD network, we introduce noise injection into the network during training, which is removed during inference.
Specifically, a noise vector $\mathbf{n}_t$ is added to the output of the $\left(t-1\right)$-th layer, serving as the input to the $t$-th layer.
The layer-wise update is expressed as
 \begin{equation}
     \mathbf{h}_t = \phi_t\left(\mathbf{h}_{t-1}+ \mathbf{n}_t; \mathbf{W}_t\right),
 \end{equation}
where $\phi_t(\cdot; \mathbf{W}_t)$ corresponds to the $t$-th layer of the unrolled PGD network, parameterized by $\mathbf{W}_t$. 
The injected noise $\mathbf{n}_t$ is drawn from $\mathcal{N}\left(0, \sigma_t^2 \mathbf{I}\right)$, where $\sigma_t^2$ decreases across the layers.
Introducing noise during training encourages each layer to perform updates with reduced reliance on the outputs from previous layers.
% The injected noise during training enforces layer-wise robustness by reducing dependence on preceding layers.
\subsection{Primal-Dual Training }
To solve the constrained learning problem in \eqref{CO}, we employ the Lagrangian dual approach.
The Lagrangian function is defined as
\begin{equation}
\mathcal{L}(\mathbf{W}, \boldsymbol{\lambda})= \mathbb{E}\left[\ell\left(\boldsymbol{\Phi}(\mathbf{y} ;  \mathbf{W}), \mathbf{h}\right)\right]+\sum_{t=1}^T \lambda_t \mathbb{E}\left[\mathcal{C}\left(\mathbf{h}_t, \mathbf{h}_{t-1}\right)\right],
\label{DO}
\end{equation}
where $\boldsymbol{\lambda} \in \mathbb{R}_{+}^T$ is a vector of dual variables, with each $\lambda_t$ associated with the monotonic descent constraint.
As the expectation $\mathbb{E}[\cdot]$ over an unknown distribution is intractable, the Lagrangian function can be approximated using empirical expectations. Specifically, we replace the expectation with the sample mean, denoted by $\widehat{\mathbb{E}}[\cdot]$, over the training data. The resulting empirical Lagrangian function is given by
\begin{equation}
\widehat{\mathcal{L}}(\mathbf{W}, \boldsymbol{\lambda})= \widehat{\mathbb{E}}\left[\ell\left(\boldsymbol{\Phi}(\mathbf{y} ; \mathbf{W}), \mathbf{h}\right)\right]+\sum_{t=1}^T \lambda_t \widehat{\mathbb{E}}\left[\mathcal{C}\left(\mathbf{h}_t, \mathbf{h}_{t-1}\right)\right].
\label{ELa}
\end{equation}
With the empirical approximation, the empirical dual problem is given as follows
\begin{equation}
\widehat{D}=\max _{\boldsymbol{\lambda} \in \mathbb{R}_{+}^T} \min _{\mathbf{W}} \widehat{\mathcal{L}}(\mathbf{W}, \boldsymbol{\lambda}).
\label{EDO}
\end{equation}

The empirical dual problem can be solved using a primal-dual optimization approach, where the optimization is conducted through alternating updates of primal and dual variables. 
The primal update minimizes the empirical Lagrangian $\widehat{\mathcal{L}}$ with respect to the network parameters $\mathbf{W}$ using gradient descent with step size $\mu_{{w}}$, as shown in \eqref{23}. 
The dual update maximizes $\widehat{\mathcal{L}}$ with respect to the dual variables $\boldsymbol{\lambda}$ using gradient ascent with step size $\mu_{\lambda}$, while ensuring non-negativity through a projection operation, as shown in \eqref{24}. 
The optimization alternates between these two updates over multiple epochs, gradually refining both the primal and dual variables until convergence, which has been theoretically established in \cite{9813433}.
The complete procedure is outlined in Algorithm~\ref{alg:primal-dual}.
\begin{algorithm}
\caption{Primal-Dual Training Algorithm}
\label{alg:primal-dual}
\begin{algorithmic}[1]
\STATE Initialize $\mathbf{W} = \{\mathbf{W}_t\}_{t=1}^T$ and $\boldsymbol{\lambda} = \{\lambda_t\}_{t=1}^T$
\FOR{each epoch}
    \FOR{each batch}
        \STATE Compute the empirical Lagrangian $\widehat{\mathcal{L}}(\mathbf{W}, \boldsymbol{\lambda})$
        \STATE Update the primal variable:
        \begin{align}
            \mathbf{W} \leftarrow \mathbf{W} - \mu_w \nabla_{\mathbf{W}}\widehat{\mathcal{L}}(\mathbf{W}, \boldsymbol{\lambda})
            \label{23}
        \end{align}
        \STATE Update the dual variable:
        \begin{align}
            \boldsymbol{\lambda} \leftarrow [\boldsymbol{\lambda} + \mu_{\lambda} \nabla_{\boldsymbol{\lambda}}\widehat{\mathcal{L}}(\mathbf{W}, \boldsymbol{\lambda})]_{+}
            \label{24}
        \end{align}
\ENDFOR
\ENDFOR
\RETURN $\mathbf{W}$
\end{algorithmic}
\end{algorithm}

Algorithm~\ref{alg:primal-dual} provides an effective approach to solving the dual problem in \eqref{EDO}. However, it is important to note that the dual problem is not strictly equivalent to the primal problem in \eqref{CO}. The difference stems from two sources: the approximation gap, arising from the dual formulation, and the estimation gap, introduced by replacing statistical expectations with empirical averages. To analyze and quantify these gaps, we leverage insights from the constrained learning theory (CLT) \cite{9813433, 10741959} under the following assumptions:
\begin{assumption} \label{assum:lipschitz}
The functions $\ell(\cdot, \mathbf{h})$ and $\mathcal{C}(\cdot, \mathbf{h})$ are bounded and satisfy $I$-Lipschitz continuity for all $\mathbf{h}$.
\end{assumption}

\begin{assumption} \label{assum:conv}
Consider $\phi_t \circ \ldots \circ \phi_1 \in \mathcal{P}_t$, which represents the composition of $t$ unrolled layers, and let $\overline{\mathcal{P}}_t = \overline{\operatorname{conv}}\left(\mathcal{P}_t\right)$ denote the convex hull of $\mathcal{P}_t$. 
Then, for each $\bar{\phi}_t \circ \ldots \circ \bar{\phi}_1 \in \overline{\mathcal{P}}_t$ and  $\nu > 0$, there exists a set of parameters $\mathbf{W}_{1:t}$ such that the following holds
$\mathbb{E}\left[\left|\phi_{t} \circ \ldots \circ\phi_1\left(\mathbf{x} ; \mathbf{W}_{1: t}\right)-\bar{\phi}_t \circ \ldots \circ\bar{\phi}_1(\mathbf{x})\right|\right] \leq \nu.$
\end{assumption}
\begin{assumption} \label{assum:condition}
Let $\mathcal{H}$ be a compact set, and suppose the conditional distribution $p\left(\mathbf{y}|\mathbf{h}\right)$ is non-atomic for every $\mathbf{h}$.
For $\boldsymbol{\Phi} \in \overline{\mathcal{P}}_t$ with $t \leq T$, the mappings $\mathbf{h} \rightarrow \ell(\boldsymbol{\Phi}(\cdot), \mathbf{h}) p(\cdot | \mathbf{h})$ and $\mathbf{h} \rightarrow \mathcal{C}(\boldsymbol{\Phi}(\cdot), \mathbf{h}) p(\cdot | \mathbf{h})$ satisfy uniform continuity in the total variation topology.
\end{assumption}
\begin{assumption}\label{assum:4}
Let $\zeta(J, \delta) \geq 0$ be a function that decreases monotonically with the number of samples $J$. With probability $1-\delta$, the following inequalities hold
$
|\mathbb{E}[\ell(\Phi(\mathbf{y}; \mathbf{W}), \mathbf{h})] - \widehat{\mathbb{E}}[\ell(\Phi(\mathbf{y}; \mathbf{W}), \mathbf{h})]| \leq \zeta(J, \delta)
$
and
$
|\mathbb{E}[\mathcal{C}(\mathbf{h}_t, \mathbf{h}_{t-1})] - \widehat{\mathbb{E}}[\mathcal{C}(\mathbf{h}_t, \mathbf{h}_{t-1})]| \leq \zeta(J, \delta)
$
for all $ t \leq T$.
\end{assumption}
\begin{assumption}\label{assum:5}
    Suppose $\boldsymbol{\Phi} \in {\mathcal{P}}_T$  is strictly feasible, so that the following conditions hold for all $ t \leq T$:
    $
    \mathbb{E}[\mathcal{C}(\mathbf{h}_t, \mathbf{h}_{t-1})] \leq -I \nu - \xi$
    and
    $\widehat{\mathbb{E}}[\mathcal{C}(\mathbf{h}_t, \mathbf{h}_{t-1})] \leq -\xi,
    $
    with $\xi > 0$.
\end{assumption}

Based on these assumptions, the CLT \cite{9813433} establishes that a stationary point of \eqref{EDO} corresponds to a solution that is probably approximately correct for the primal problem in \eqref{CO}.
\begin{theorem}
Consider $\left(\mathbf{W}^*, \boldsymbol{\lambda}^*\right)$ as a stationary point of the dual problem in \eqref{EDO}. Under Assumptions 1-5, it holds that
\begin{align}
&|P-\widehat{D}| \leq I \nu+\gamma \zeta(J, \delta); \label{dualgap}\\
&\mathbb{E}\left[\mathcal{C}\left(\mathbf{h}_t, \mathbf{h}_{t-1}\right)\right] \leq \zeta(J, \delta), \forall t,\label{satcon}
\end{align}
with probability $1 - \delta$. The constant $\gamma$ satisfies $\gamma \geq \max \left\{\|\boldsymbol{\lambda}^*\right\|,\|\overline{\boldsymbol{\lambda}}^*\|\}$, where $\overline{\boldsymbol{\lambda}}^*=\operatorname{argmax}_{\boldsymbol{\lambda}} \min _{\mathbf{W}} \mathcal{L}(\mathbf{W}, \boldsymbol{\lambda})$.
\end{theorem}

According to Theorem 1, the duality gap depends on the constant $I$, the richness parameter $\nu$, the sample complexity $\zeta(J, \delta)$, and the constant $\gamma$.
The constant $\gamma$ is determined by the dual variables and reflects the sensitivity of the constraints.
Moreover, it shows that the descending constraints in each layer are satisfied with a probability of $1 - \delta$ and are subject to a bound. 
This bound, determined by the sample complexity, can be reduced by increasing the number of samples $J$.
\subsection{Convergence}
To analyze the convergence behavior of the constrained unrolled PGD network, we investigate how the sequence of outputs generated by the network gradually approaches the optimal solution.
\begin{theorem}{\label{theorem2}}
Suppose there exists a near-optimal solution to the constrained optimization problem in \eqref{CO} that satisfies the monotonic descent constraint in \eqref{cons} with a probability of $1-\delta$. Under Assumptions 1–5, the sequence of outputs exhibits the following convergence behavior
\begin{equation}
\lim _{t \rightarrow \infty}  \mathbb{E}\left[\min _{k \leq t} \left\|\mathbf{h}_k-\mathbf{h}\right\|_2 \right] \leq \frac{1}{\epsilon}\left(\zeta(J, \delta)+\frac{\delta C}{1-\delta}\right), \quad \text { a.s. } 
\end{equation}
\end{theorem}

Theorem \ref{theorem2} characterizes the convergence behavior of the constrained unrolled PGD network.
It demonstrates that the sequence of outputs infinitely often enters a region around the true solution $\mathbf{h}$.
In this region,  the expected value of the distance norm is bounded above by $\frac{1}{\epsilon}\left(\zeta(J, \delta)+\frac{\delta C}{1-\delta}\right)$.
The characteristics of this region are governed by the sample complexity of $\zeta(J, \delta)$, the bound  constant $C$, and a design parameter $\epsilon$ of the constraints.
The detailed proof of Theorem \ref{theorem2} is provided in Appendix A.
\section{Simulation Results}

\subsection{Simulation Setup}
Building upon the system model described in Section II, the following parameters are adopted for performance evaluation.
The system employs $M=256$ subcarriers operating at a center frequency of $f_c = 100~\mathrm{GHz}$. 
Each subcarrier is assigned a bandwidth of $B=10~\mathrm{GHz}$.
The BS is equipped with $N_{\mathrm{RF}}=4$ RF chains, and the combining matrix $\mathbf{A}$ has entries that are randomly selected from the set $\{\pm 1/\sqrt{N}\}$.

The channel parameters for the ULA and UPA configurations are specified as follows.
For the ULA configuration, the BS is equipped with $N=512$ antenna elements spaced at half-wavelength intervals $d=\lambda/2$. 
The channel comprises $L=3$ multipath components, with the AoA $\theta_l$ for each path randomly sampled from the interval $[-\pi/3, \pi/3]$. 
For the UPA configuration, the BS employs $N_1 \times N_2 = 256 \times 8$ elements, totaling $N=2048$ antenna elements with half-wavelength spacing.
The channel similarly comprises $L=3$ multipath components, where both the azimuth angle $\phi_l$ and elevation angle $\theta_l$ for each path are randomly sampled from $[-\pi/6, \pi/6]$. 
For both array configurations, the complex path gains $\alpha_{l,m}$ are drawn from $\mathcal{CN}(0,1)$.
In addition, the distance $r_l$ between each scatterer and the array center is randomly sampled from the range of $5$ m to $30$ m.
To model the spatial non-stationarity, the visibility region for each path is determined by randomly generating a discrete coverage rate and selecting a corresponding contiguous block from the array. For instance, in a ULA with $N$ antennas, the coverage rate $\alpha$ is chosen from a discrete set such as $\{0.25, 0.5, 0.75\}$, and a contiguous block of $\lceil \alpha N \rceil$ antennas is selected as the visible region for that path. A similar approach is applied for the UPA by randomly selecting a rectangular block corresponding to the chosen coverage rate.
% To model the spatial non-stationarity, the visibility region $\mathrm{VR}_l$ for each path is implemented by dividing the array into subarrays and randomly assigning visibility to these subarrays, with at least one subarray guaranteed to be visible. Specifically, the ULA is divided into $4$ equal subarrays, while the UPA is partitioned into an $4 \times 2$ grid of subarrays.

The generated channel data are used to create data pairs $\{\mathbf{y}_m, \mathbf{h}_m\}$ for signal-to-noise ratio (SNR) levels ranging from $-5$ dB to $10$ dB.
For each SNR level, $5 \times 10^7$ data pairs are generated and divided into training, validation, and test datasets with $3.5 \times 10^7$, $0.5 \times 10^7$, and $1.0 \times 10^7$ pairs, respectively.
The proposed network consists of $T = 5$ layers and is implemented in PyTorch. 
It is trained on a single NVIDIA RTX 3090 GPU using the Adam optimizer with a learning rate of $1 \times 10^{-4}$.
The data pairs across all $256$ subcarriers are stacked to form a batch, aligning with the system's subcarrier configuration and enabling parallel processing on the GPU.
 
To demonstrate the effectiveness of the proposed constrained unrolled PGD network, its performance is compared with the following benchmark algorithms:
\begin{itemize}
    \item {LMMSE}
    : The linear minimum mean squared error (LMMSE) estimator exploits the prior statistical characteristics of the channel to construct a linear estimator. 
    \item {OMP}: The orthogonal matching pursuit algorithm employs polar-domain dictionaries\cite{9693928,10123941} for sparse channel reconstruction.
    \item {ISTA-Net+}: A deep unrolling network inspired by the iterative shrinkage-thresholding algorithm (ISTA)\cite{Zhang_2018_CVPR}.
    \item{AMP-SBL}: A deep unrolling network combining approximate message passing and sparse Bayesian learning with learned parameters for channel estimation\cite{gao2024deep}.
    \item{D2-CNN}: A data-driven CNN utilizing the same U-Net architectural design as the proximal step network $\mathcal{N}_{\theta_t}$  in the proposed algorithm, but trained end-to-end to directly map the received signal to the channel estimate.
    
\end{itemize}
Table~\ref{tab:method_comparison} summarizes the computational complexity of deep learning-based methods, including ISTA-Net+, AMP-SBL, D2-CNN, and the proposed constrained unrolled PGD network, evaluated under the ULA configuration.
\begin{table}[htbp]
\caption{Computational Complexity}
\label{tab:method_comparison}
\centering
\begin{tabular}{|l|c|}
\hline
\textbf{Method} & \textbf{FLOPs ($\times 10^9$)} \\
\hline
ISTA-Net+ & 2.54 \\
AMP-SBL  & 2.16 \\
D2-CNN & 3.77 \\
PGD-Net   & 3.79 \\
\hline
\end{tabular}
\end{table}
%PGD-3.79G

For quantitative evaluation of channel estimation accuracy, the normalized mean squared error (NMSE) is employed as the performance metric, defined as
\begin{equation}
    \text{NMSE} = \mathbb{E}\left[ \frac{\|\mathbf{h} - \hat{\mathbf{h}}\|_2^2}{\|\mathbf{h}\|_2^2} \right],
\end{equation}
where $\mathbf{h}$ denotes the true channel vector, and $\hat{\mathbf{h}}$ represents its corresponding estimate obtained by a given algorithm.

\subsection{Performance Evaluation}
\begin{figure}[htbp]
    \centering
    \begin{subfigure}{\columnwidth}
        \centering
        \includegraphics[width=\columnwidth]{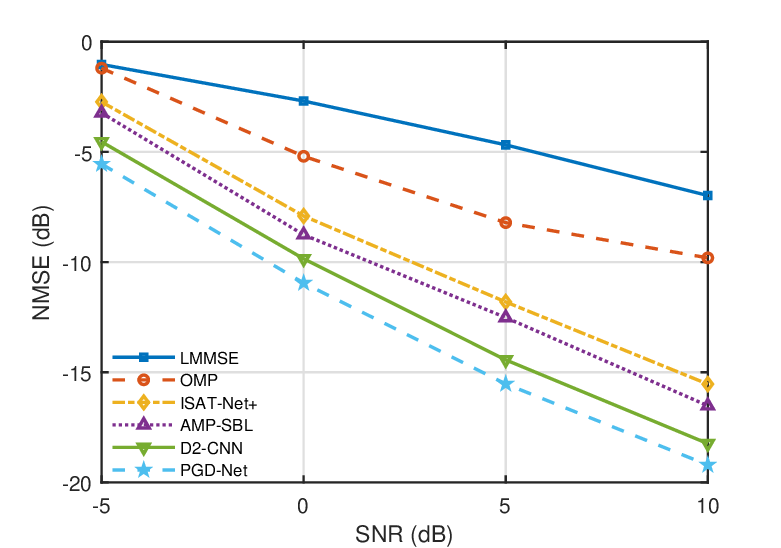}
        \caption{NMSE versus SNR under ULA configuration}
        \label{ULA_SNR}
    \end{subfigure}
    \vspace{0.5em}
    \begin{subfigure}{\columnwidth}
        \centering
        \includegraphics[width=\columnwidth]{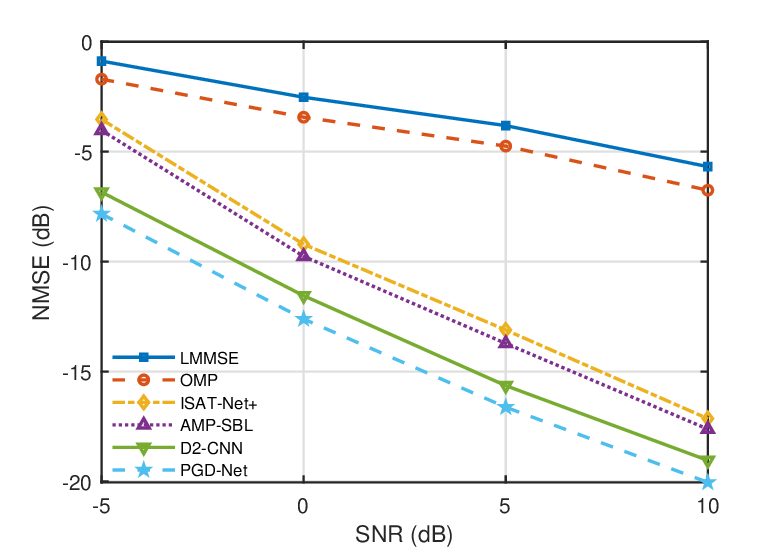}
        \caption{NMSE versus SNR under UPA configuration}
        \label{UPA_SNR}
    \end{subfigure}
    \caption{NMSE performance versus SNR under ULA and UPA configurations.}
    \label{fig:SNR_compare}
\end{figure}

The performance is first evaluated under the ULA configuration with a pilot overhead of 256.
As observed in Fig. \ref{ULA_SNR}, traditional methods such as LMMSE and OMP demonstrate limited adaptability to complex channel characteristics, which results in relatively poorer channel estimation accuracy.
Deep unrolling-based algorithms, including ISTA-Net+ and AMP-SBL, demonstrate enhanced performance compared to  the traditional methods, leveraging learned parameters to better capture channel structure.
Nonetheless, their performance remains inferior to the proposed constrained unrolled PGD network, suggesting that the adopted architecture and the learned proximal operator enhance the model's capability to implicitly capture intricate channel characteristics.
Furthermore, despite employing a similar U-Net architecture, the D2-CNN approach achieves lower channel estimation accuracy than the proposed PGD network.
This demonstrates the effectiveness of incorporating iterative optimization into neural networks, enabling the integration of model-based information and data-driven adaptability.

Under the UPA configuration with a pilot overhead of 1024, a performance trend similar to that in the ULA scenario is clearly observed in Fig. \ref{UPA_SNR}.
Specifically, at an SNR of $0$ dB, the proposed method achieves an NMSE of around $-12.61$ dB, outperforming the AMP-SBL and ISTA-Net+ methods by approximately $2.85$ dB and $3.39$ dB, respectively.
When compared with the conventional LMMSE and OMP algorithms, the performance improvement increases significantly to about $10.08$ dB and $9.17$ dB, respectively.
Furthermore, at an SNR of $10$ dB, the NMSE gap widens significantly, with the proposed PGD network achieving an NMSE improvement of more than $14.36$ dB over LMMSE and nearly $13.28$ dB over the OMP algorithm.
It is also worth noting that traditional algorithms exhibit diminishing performance as the antenna dimensionality increases. 
For instance, the LMMSE achieves an NMSE of approximately $-6.98$ dB at an SNR of $10$ dB in the ULA scenario with 512 antennas, but its performance deteriorates slightly to around $-5.67$ dB when employing the UPA with 2048 antennas. 
In contrast, deep learning-based approaches, such as our constrained unrolled PGD network, benefit from increased dimensionality.
Specifically, the proposed method attains an NMSE of roughly $-19.21$ dB at an SNR of $10$ dB in the ULA case, and further improves to nearly $-20.04$ dB in the UPA configuration.
This performance enhancement can be attributed to the capability of deep neural networks to effectively exploit and capture the high-dimensional and complex channel structure in XL-MIMO systems.

\begin{figure}[htbp]
    \centering
    \begin{subfigure}{\columnwidth}
        \centering
        \includegraphics[width=\columnwidth]{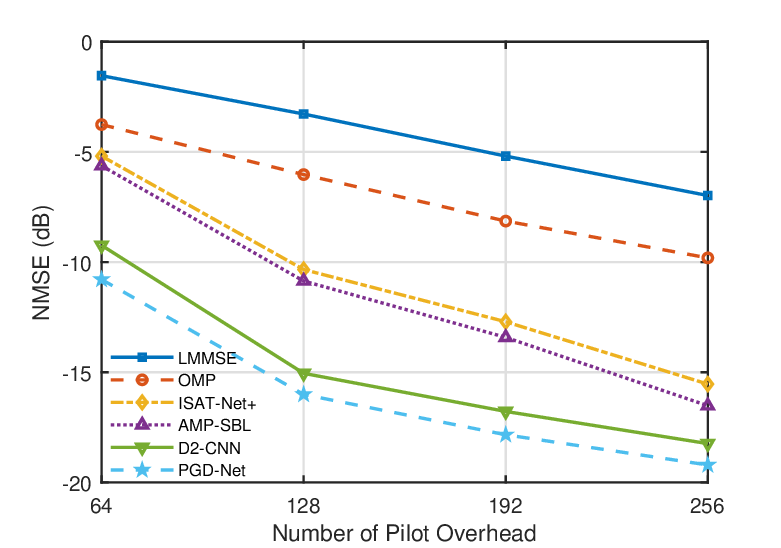}
        \caption{NMSE versus pilot overhead under ULA configuration}
        \label{ULA_PILOT}
    \end{subfigure}
    \vspace{0.5em}
    \begin{subfigure}{\columnwidth}
        \centering
        \includegraphics[width=\columnwidth]{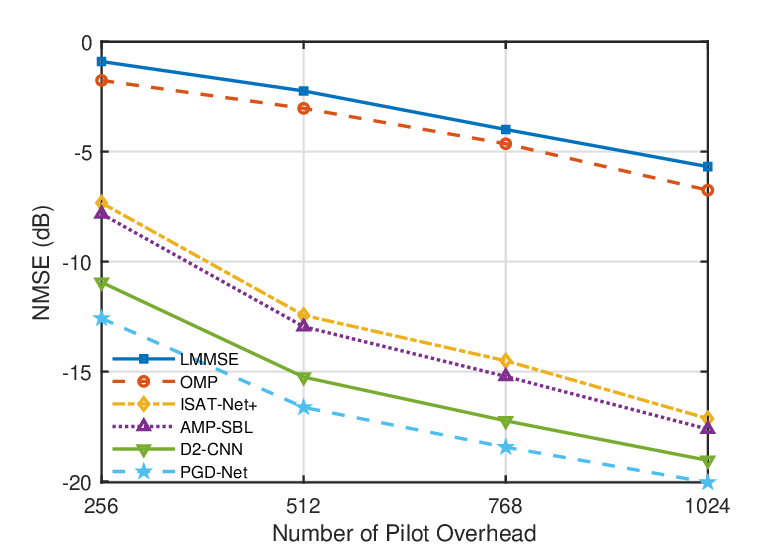}
        \caption{NMSE versus pilot overhead under UPA configuration}
        \label{UPA_PILOT}
    \end{subfigure}
    \caption{NMSE performance versus pilot overhead under ULA and UPA configurations.}
    \label{fig:PILOT_compare}
\end{figure}

The impact of varying the number of pilot overhead on channel estimation accuracy for the ULA configuration under a fixed SNR of 10 dB is presented in Fig. \ref{ULA_PILOT}. 
It is observed that increasing the pilot overhead generally enhances estimation accuracy for all considered methods, as additional pilot overhead enriches the measurement information.
Compared with conventional algorithms (e.g., LMMSE and OMP) and other deep learning-based approaches (e.g., ISTA-Net+, AMP-SBL, and D2-CNN), the proposed constrained unrolled PGD network consistently demonstrates superior estimation performance across all pilot overhead settings.
%Although conventional methods also benefit from increased pilot overhead, their performance improvements remain relatively moderate.
%In contrast, deep learning-based methods, particularly the proposed PGD network, exhibit more significant performance gains. 
Although conventional methods also benefit from increased pilot overhead, their performance improvements remain relatively moderate.
Deep learning-based methods, especially the proposed PGD network, consistently achieve lower NMSE compared to conventional approaches.
This result demonstrates the strong capability of the deep neural network architecture to effectively utilize the enriched measurement information provided by additional pilot overhead.
The superior performance of the proposed method is primarily attributed to a data-driven proximal module, which implicitly learns underlying channel characteristics directly from data.
As the number of pilot overhead increases, this learned prior becomes increasingly accurate and informative, enabling the network to better capture inherent channel structure and thus enhance estimation accuracy.

The results for varying pilot overhead under the UPA configuration at a fixed SNR of 10 dB demonstrate similar performance trends to those observed in the ULA.
As shown in Fig. \ref{UPA_PILOT}, the proposed PGD network achieves superior performance across all pilot overhead settings. 
Specifically, at $256$ pilot overhead, the PGD-Net attains an NMSE of $-12.58$ dB, outperforming D2-CNN at $-10.94$ dB, AMP-SBL at $-7.83$ dB, ISTA-Net+ at $-7.33$ dB, LMMSE at $-0.91$ dB, and OMP at $-1.76$ dB.
When increasing pilot overhead to $512$ symbols, the PGD network achieves an NMSE of $-16.63$ dB. 
In comparison, D2-CNN, AMP-SBL, and ISTA-Net+ reach $-15.52$ dB, $-12.96$ dB, and $-12.44$ dB, respectively. 
The conventional LMMSE and OMP methods improve to $-2.23$ dB and $-3.03$ dB.
The consistent performance advantage across both ULA and UPA configurations validates the robustness and generalizability of the proposed network.

\begin{figure}[htbp]
\centering
\includegraphics[width=\columnwidth]{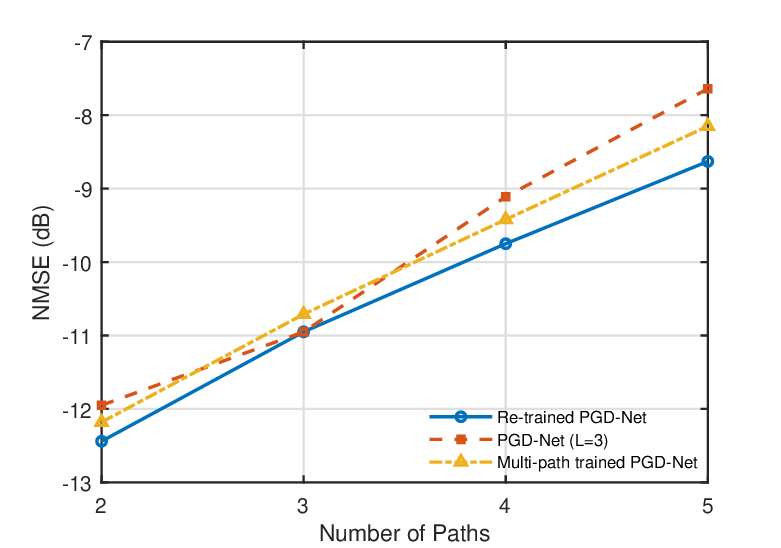}
\caption{NMSE performance versus number of paths under ULA configuration.}
\label{nmse_vs_path}
\end{figure}

Fig. \ref{nmse_vs_path} presents the NMSE performance of the proposed PGD network with varying numbers of paths under a fixed SNR of 0 dB for the ULA configuration.
Three training strategies are considered:
(1) Re-trained PGD-Net, where the network is individually trained and evaluated at each specific number of paths;
(2) PGD-Net, where the network is trained only at $L=3$ and evaluated at different path numbers;
(3) Multi-path trained PGD-Net, where the network is trained jointly using data from multiple path scenarios $(L=2,3,4,5)$.
It is observed that the NMSE performance of all three strategies consistently deteriorates as the number of paths increases, demonstrating that channel estimation becomes increasingly challenging as the number of paths increases.
Specifically, the re-trained PGD serves as a performance benchmark, achieving the lowest NMSE values across all considered scenarios due to its dedicated training at each specific path number.
Compared with the re-trained PGD baseline, the PGD-Net maintains robust estimation capability, achieving an NMSE reduction of approximately $0.49$ dB at $L=2$, but experiencing a degradation of about $0.98$ dB at $L=5$.
This observation demonstrates the model's generalization ability to scenarios with fewer paths while sustaining reasonable performance for higher numbers of paths, even without specialized training.
Furthermore, the multi-path trained PGD-Net exhibits enhanced generalization capability and robustness compared to the PGD-Net, achieving approximately $0.48$ dB NMSE improvement at $L=5$.

\begin{figure}[htbp]
\centering
\includegraphics[width=\columnwidth]{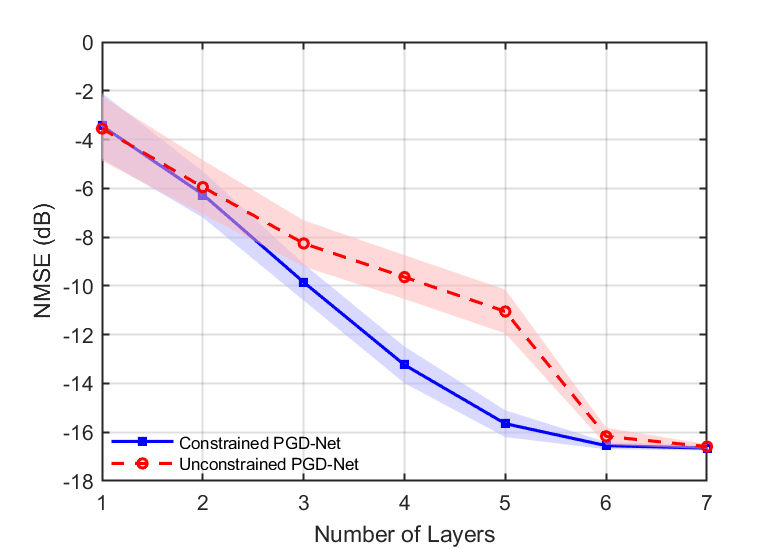}
\caption{NMSE performance versus number of unrolled layers.}
\label{conver}
\end{figure}

Fig. \ref{conver} illustrates the NMSE performance across different unrolled layers at a fixed SNR of 5 dB under the UPA configuration, comparing the proposed constrained PGD-Net with its unconstrained counterpart.
It can be observed that both constrained and unconstrained PGD-Net eventually converge to comparable performance.
However, their convergence behaviors differ across intermediate layers.
Specifically, the constrained PGD-Net demonstrates a monotonic decrease in NMSE as the number of layers increases, reflecting the enforced constraint that promotes convergence towards the optimal solution at each intermediate layer.
In contrast, although the unconstrained PGD-Net also exhibits an overall decreasing trend, it experiences relatively greater performance variation at intermediate layers, as indicated by the wider shaded confidence intervals.
These results validate the effectiveness of the monotonic descent constraint, which stabilizes the intermediate convergence behavior of the PGD-Net and ensures consistent improvement in performance at each layer.
\section{Conclusion}
In this paper, we considered channel estimation in wideband XL-MIMO systems, where the channel is characterized by near-field propagation, beam split, and spatial non-stationarity.
To address the challenges posed by these channel characteristics, we formulated channel estimation as a MAP problem and proposed an unrolled PGD network with learnable step sizes. 
The proposed network employs a neural network for proximal mapping, enabling the implicit learning of prior channel knowledge without the need for explicit regularization functions.
To enhance the convergence, we introduced a monotonic descent constraint on the layer-wise estimation error and developed a primal-dual training algorithm.
Theoretical analyses were conducted to characterize the network's convergence behavior and simulation results validated the network's effectiveness.
%Theoretical analysis established results on both the duality gap and the convergence behavior of the approach, and simulation results validated its effectiveness.
% In this paper, we addressed the channel estimation in wideband XL-MIMO systems characterized by near-field propagation, frequency-dependent beam split, and spatial non-stationary effects. 
% We formulated channel estimation as a MAP problem and proposed an unrolled PGD network architecture. 
% By integrating learnable step sizes and replacing the proximal operator with a neural network, our approach effectively captures channel characteristics without requiring explicit regularization terms.
% To enhance the performance of unrolled PGD(To further enhance the robustness of the unroll PGD), we incorporated a monotonic descent constraint（具体一点） , solving the resulting constrained optimization problem with a primal-dual approach.
% Our theoretical analysis provided guarantees on both the duality gap and the convergence behavior of our approach.
% 实验结果验证了我们的方法的有效性。

% if have a single appendix:
%\appendix[Proof of the Zonklar Equations]
% or
%\appendix  % for no appendix heading
% do not use \section anymore after \appendix, only \section*
% is possibly needed

% use appendices with more than one appendix
% then use \section to start each appendix
% you must declare a \section before using any
% \subsection or using \label (\appendices by itself
% starts a section numbered zero.)
%

\appendices
\section{Proof of Convergence}
Let $(\Omega, \mathcal{F}, P)$ denote a probability space, where $\Omega$ represents the sample space, $\mathcal{F}$ denotes a sigma-algebra, and $P:\mathcal{F} \rightarrow[0,1]$ represents a probability measure.
For a random variable $X : \Omega \to \mathbb{R}$, we write $P(X = 0)$ instead of $P({\omega : X(\omega) = 0})$ for notational simplicity.
Additionally, let $\left\{\mathcal{F}_t\right\}_{t>0}$  denote a filtration of $\mathcal{F}$, which represents an increasing sequence of sigma-algebras satisfying $\mathcal{F}_{t-1} \subset \mathcal{F}_t$. 
The outputs of the unrolled layers $\mathbf{h}_t$  are assumed to be adapted to this filtration, i.e., $\mathbf{h}_t \in \mathcal{F}_t$.

The proof of Theorem 2 is outlined in the following.
By letting $A_t \in \mathcal{F}_t$ denote the event that the constraint in \eqref{satcon} is satisfied, and applying the total expectation theorem, we obtain
\begin{equation}
\begin{aligned}
\mathbb{E}\left[\left\|\mathbf{h}_t-\mathbf{h}\right\|_2\right]= & P\left(A_t\right) \mathbb{E}\left[\left\|\mathbf{h}_t-\mathbf{h}\right\|_2 \mid A_t\right] \\
+&P\left(A_t^c\right) \mathbb{E}\left[\left\|\mathbf{h}_t-\mathbf{h}\right\|_2 \mid A_t^c\right],
\end{aligned}
\label{TE}
\end{equation}
where $P\left(A_t\right)=1-\delta$.
The first term represents the conditional expectation under the event that the constraint is satisfied, and it is bounded as
\begin{equation}
    \mathbb{E}\left[\left\|\mathbf{h}_t-\mathbf{h}\right\|_2 \mid A_t\right] \leq (1-{\epsilon})\mathbb{E}\left[\left\|\mathbf{h}_{t-1}-\mathbf{h}\right\|_2\right] + \zeta(J, \delta).
\end{equation}
The second term corresponds to the complementary event $A_t^c \in \mathcal{F}_t$, where the conditional expectation is bounded by the maximum value of the random variable, satisfying
\begin{equation}
  \mathbb{E}\left[\left\|\mathbf{h}_t-\mathbf{h}\right\|_2 \mid A_{l}^c\right] \leq \max _{\mathbf{h}_t}\left\|\mathbf{h}_t-\mathbf{h}\right\|_2 \leq C,
\end{equation}
where $C$ is the bound implied by Assumption 1.
Combining these results, the total expectation in \eqref{TE} can be bounded as
\begin{equation}
\begin{aligned}
\mathbb{E}\left[\left\|\mathbf{h}_t-\mathbf{h}\right\|_2\right] \leq& (1-\delta)(1-\epsilon)\mathbb{E}\left[\left\|\mathbf{h}_{t-1}-\mathbf{h}\right\|_2\right]\\
&+(1-\delta) \zeta(J, \delta)+\delta C,
\end{aligned}
\label{ineq}
\end{equation}
which holds almost surely.

We define \(Z_t = \mathbb{E}_{D_y}\left[\|\mathbf{h}_t - \mathbf{h}\|_2\right]\) as the expected distance norm, averaged over the input data distribution \(D_y\), which depends on \(\mathbf{h}_{t-1}\) and the noise \(\mathbf{n}_t\).
As rigorously established in \cite{10741959}, the following holds almost surely
\begin{equation}
\liminf _{t \rightarrow \infty}\left(\epsilon Z_t-\zeta(J, \delta)-\frac{\delta C}{1-\delta}\right)=0.
\end{equation}

To establish convergence, we start by noting that $Z_t=\int\left\|\mathbf{h}_t-\mathbf{h}\right\|_2 dP$, which allows us to rewrite the result as
\begin{equation}
\liminf _{t \rightarrow \infty} \int \epsilon\left\|\mathbf{h}_t-\mathbf{h}\right\|_2 d P=\zeta(J, \delta)+\frac{\delta C}{1-\delta}, \quad \text { a.s. }
\end{equation}
By applying Fatou's lemma, we obtain the following inequality
\begin{equation}
\int \liminf _{t \rightarrow \infty} \epsilon\left\|\mathbf{h}_t-\mathbf{h}\right\|_2 d P \leq \liminf _{t \rightarrow \infty} \int \epsilon\left\|\mathbf{h}_t-\mathbf{h}\right\|_2 d P.
\end{equation}
It follows that, for sufficiently large $t$, the relationship
\begin{equation}
    \min _{k \leq t} \left\|\mathbf{h}_k-\mathbf{h}\right\|_2\leq \liminf _{t \rightarrow \infty} \left\|\mathbf{h}_t-\mathbf{h}\right\|_2
\end{equation}
holds, which leads to
\begin{equation}
  \epsilon \int \min _{k \leq t} \left\|\mathbf{h}_k-\mathbf{h}\right\|_2 d P \leq \zeta(J, \delta)+\frac{\delta C}{1-\delta}, \quad \text { a.s. }
\end{equation}
Taking the limit as $t \rightarrow \infty$, the resulting inequality is
\begin{equation}
\lim _{t \rightarrow \infty} \int \min _{k \leq t} \left\|\mathbf{h}_k-\mathbf{h}\right\|_2 d P \leq \frac{1}{\epsilon}\left(\zeta(J, \delta)+\frac{\delta C}{1-\delta}\right), \quad \text { a.s. }
\end{equation}
Finally, the expectation satisfies
\begin{equation}
\lim _{t \rightarrow \infty}  \mathbb{E}\left[\min _{k \leq t} \left\|\mathbf{h}_k-\mathbf{h}\right\|_2 \right] \leq \frac{1}{\epsilon}\left(\zeta(J, \delta)+\frac{\delta C}{1-\delta}\right), \quad \text { a.s. } 
\end{equation}

%\section{}      
%Appendix two text goes here.

% use section* for acknowledgment
%\section*{Acknowledgment}

%The authors would like to thank...

% Can use something like this to put references on a page
% by themselves when using endfloat and the captionsoff option.
\ifCLASSOPTIONcaptionsoff
  \newpage
\fi

% trigger a \newpage just before the given reference
% number - used to balance the columns on the last page
% adjust value as needed - may need to be readjusted if
% the document is modified later
%\IEEEtriggeratref{8}
% The "triggered" command can be changed if desired:
%\IEEEtriggercmd{\enlargethispage{-5in}}

% references section

% can use a bibliography generated by BibTeX as a .bbl file
% BibTeX documentation can be easily obtained at:
% http://mirror.ctan.org/biblio/bibtex/contrib/doc/
% The IEEEtran BibTeX style support page is at:
% http://www.michaelshell.org/tex/ieeetran/bibtex/
%\bibliographystyle{IEEEtran}
% argument is your BibTeX string definitions and bibliography database(s)
%\bibliography{IEEEabrv,../bib/paper}
%
% <OR> manually copy in the resultant .bbl file
% set second argument of \begin to the number of references
% (used to reserve space for the reference number labels box)
% \begin{thebibliography}{1}

\bibliographystyle{IEEEtran}
\bibliography{bare_jrnl}
% biography section
% 
% If you have an EPS/PDF photo (graphicx package needed) extra braces are
% needed around the contents of the optional argument to biography to prevent
% the LaTeX parser from getting confused when it sees the complicated
% \includegraphics command within an optional argument. (You could create
% your own custom macro containing the \includegraphics command to make things
% simpler here.)
%\begin{IEEEbiography}[{\includegraphics[width=1in,height=1.25in,clip,keepaspectratio]{mshell}}]{Michael Shell}
% or if you just want to reserve a space for a photo:

%\begin{IEEEbiography}{Michael Shell}
%Biography text here.
%\end{IEEEbiography}

% if you will not have a photo at all:
%\begin{IEEEbiographynophoto}{John Doe}
%Biography text here.
%\end{IEEEbiographynophoto}

% insert where needed to balance the two columns on the last page with
% biographies
%\newpage

%\begin{IEEEbiographynophoto}{Jane Doe}
%Biography text here.
%\end{IEEEbiographynophoto}

% You can push biographies down or up by placing
% a \vfill before or after them. The appropriate
% use of \vfill depends on what kind of text is
% on the last page and whether or not the columns
% are being equalized.

%\vfill

% Can be used to pull up biographies so that the bottom of the last one
% is flush with the other column.
%\enlargethispage{-5in}

% that's all folks
\end{document}